\documentclass[12pt,onecolumn]{IEEEtran}

\usepackage{amssymb}
\usepackage{graphicx}
\usepackage{amsmath}
\usepackage[colorlinks,linkcolor=blue,citecolor=red,linktocpage=true]{hyperref}

\begin {document}


\def\bbbf{{\rm I\!F}}

\def\bbbz{{\mathchoice {\hbox{$\sf\textstyle Z\kern-0.4em Z$}}
{\hbox{$\sf\textstyle Z\kern-0.4em Z$}}
{\hbox{$\sf\scriptstyle Z\kern-0.3em Z$}}
{\hbox{$\sf\scriptscriptstyle Z\kern-0.2em Z$}}}}

\newtheorem{definition}{Definition}

\newtheorem{theorem}{Theorem}
\newtheorem{lemma}{Lemma}
\newtheorem{corollary}{Corollary}
\newtheorem{remark}{Remark}
\newtheorem{example}{Example}


\title{An extended characterization of a class of optimal three-weight cyclic codes over any finite field}

\author{
Gerardo Vega\thanks{G. Vega is with the Direcci\'on General de C\'omputo y de Tecnolog\'{\i}as de Informaci\'on y Comunicaci\'on, Uni\-ver\-si\-dad Nacional Aut\'onoma de 
M\'exico, 04510 Ciudad de M\'exico, MEXICO (e-mail: gerardov@unam.mx).}\thanks{Manuscript partially supported by PAPIIT-UNAM IN107515.}}
\maketitle


\begin{abstract} 
A characterization of a class of optimal three-weight cyclic codes of dimension 3 over any finite field was recently presented in \cite{Vega3}. Shortly after this, a generalization for the sufficient numerical conditions of such characterization was given in \cite{Heng}. The main purpose of this work is to show that the numerical conditions found in \cite{Heng}, are also necessary. As we will see later, an interesting feature of the present work, in clear contrast with these two preceding works, is that we use some new and non-conventional methods in order to achieve our goals. In fact, through these non-conventional methods, we not only were able to extend the characterization in \cite{Vega3}, but also present a less complex proof of such extended characterization, which avoids the use of some of the sophisticated --but at the same time complex-- theorems, that are the key arguments of the proofs given in \cite{Vega3} and \cite{Heng}. Furthermore, we also find the parameters for the dual code of any cyclic code in our extended characterization class. In fact, after the analysis of some examples, it seems that such dual codes always have the same parameters as the best known linear codes.
\end{abstract}


\section{Introduction}\label{secuno}
In coding theory an interesting but at the same time a difficult problem is to determine the weight distribution of a given code. The weight distribution is important because it plays a significant role in determining the capabilities of error detection and correction of a code. For cyclic codes this problem is even more important because this kind of codes possess a rich algebraic structure. On the other hand, it is known that cyclic codes with few weights have a great practical importance in coding theory and cryptography, and this is so because they are useful in the design of frequency hopping sequences and in the development of secret sharing schemes. A characterization of a class of optimal three-weight cyclic codes of dimension 3, over any finite field, was recently presented in \cite{Vega3}, and almost immediately after this, a generalization for the sufficient numerical conditions of such characterization was given in \cite{Heng}. By means of this generalization it was found a class of optimal three-weight cyclic codes of dimension greater than or equal to 3 that includes the class of cyclic codes characterized in \cite{Vega3}. 

The main purpose of this work is to show that the numerical conditions that were found in \cite{Heng} are also necessary. As we will see later, an interesting feature of the present work is that, in clear contrast with \cite{Vega3} and \cite{Heng}, we use some new and non-conventional methods in order to achieve our goals. More specifically, we will use the remainder operator (see next section for a formal definition of it) as one of the key tools of this work. In fact, through this remainder operator, we not only were able to extend the characterization in \cite{Vega3}, but also present a less complex proof for such extended characterization, which avoids the use of some of the sophisticated --but at the same time complex-- theorems (for example the Davenport-Hasse Theorem), that are the key arguments of the proofs given in \cite{Vega3} and \cite{Heng}. As a consequence, we were also able to present a simplified and self-contained proof of our extended characterization. As a further result, we also find the parameters for the dual code of any cyclic code in our extended characterization class. In fact, after the analysis of some examples, it seems that such dual codes always have the same parameters as the best known linear codes.

In order to provide a detailed explanation of what are the main results of this work, let $q$ and $k$ be positive integers such that $q$ is a power of a prime number, and fix $\Delta=\frac{q^k-1}{q-1}$. Also let $\gamma$ be a fixed primitive element of $\bbbf_{q^k}$. For any integer $a$, denote by $h_a(x) \in \bbbf_{q}[x]$ the minimal polynomial of $\gamma^{-a}$. With this notation in mind, the following result gives a full description for the weight distribution of a class of optimal three-weight cyclic codes of length $q^k-1$ and dimension greater than or equal to 3.

\begin{center}
TABLE I \\
{\em Weight distribution of ${\cal C}_{(\Delta e_1,e_2)}$.}
\end{center}
\begin{center}
\begin{tabular}{|c|c|} \hline
{\bf Weight} & $\;$ {\bf Frequency} $\;$\\ \hline \hline
0 & 1 \\ \hline
$q^{k-1}(q-1)-1$ & $(q-1)(q^k-1)$ \\ \hline
$q^{k-1}(q-1)$ & $q^k-1$ \\ \hline
$q^k-1$ & $q-1$ \\ \hline 
\end{tabular}
\end{center}

\begin{theorem}\label{teouno}
Suppose $k > 1$, and for any two integers $e_1$ and $e_2$, let ${\cal C}_{(\Delta e_1,e_2)}$ be the cyclic code, over $\bbbf_{q}$, with parity-check polynomial $h_{\Delta e_1}(x)h_{e_2}(x)$. If $\gcd(q-1,ke_1-e_2)=1$ and $\gcd(\Delta,e_2)=1$ then 

\begin{enumerate}
\item[{(A)}] $\deg(h_{\Delta e_1}(x))=1$ and $\deg(h_{e_2}(x))=k$. In addition, $h_{\Delta e_1}(x)$ and $h_{e_2}(x)$ are the parity-check polynomials of two different one-weight irreducible cyclic codes of length $q^k-1$, whose nonzero weights are, respectively, $q^k-1$ and $q^{k-1}(q-1)$.
\item[{(B)}] ${\cal C}_{(\Delta e_1,e_2)}$ is an optimal three-weight $[q^k-1,k+1,q^{k-1}(q-1)-1]$ cyclic code over $\bbbf_{q}$, with the weight distribution given in Table I. In addition, if $B_j$, with $0<j\leq q^k-1$, is the number of words of weight $j$ in the dual code of ${\cal C}_{(\Delta e_1,e_2)}$, then $B_1=B_2=0$, and 

$$B_3=\frac{(q^k-3)(q^k-1)(q-2)(q-1)}{6} \; .$$

\noindent
Thus, if $q>2$, then this dual is a single-error-correcting code with parameters $[q^k-1,q^k-2-k,3]$. 
\end{enumerate}
\end{theorem}

Since the previous reducible cyclic codes are optimal, a natural question that arises is whether there exist other cyclic codes (reducible or irreducible, and apart from those in Theorem \ref{teouno}), whose weight distribution is given in Table I. That is, we ask ourselves if the numerical conditions in Theorem \ref{teouno}, are also necessary. The answer is yes, and we formally state this result in the following:

\begin{theorem}\label{teodos}
Suppose $k > 1$, and let ${\cal C}$ be a cyclic code of length $q^k-1$ over $\bbbf_{q}$. Then, the weight distribution of ${\cal C}$ is given in Table I if and only if its dimension is $k+1$, and there exist two integers, $e_1$ and $e_2$, in such a way that $h_{\Delta e_1}(x)h_{e_2}(x)$ is the parity-check polynomial of ${\cal C}$, and where these integers satisfy the two conditions: $\gcd(q-1,k e_1-e_2)=1$ and $\gcd(\Delta,e_2)=1$.
\end{theorem} 

As can be seen, the previous result is the natural extension of the characterization given in Theorem 5 of \cite{Vega3}. Now, note that the kind of characterizations that are given in terms of a weight distribution table are, in general, very difficult to establish. One of the most relevant efforts in that direction is the work of B. Schmidt and C. White in \cite{Schmidt}, where simple necessary and sufficient numerical conditions for an irreducible cyclic code to have at most two weights, are presented. As will be clear later, this important work will be essential in order to present a formal proof of the extended characterization in Theorem \ref{teodos}.

This work is organized as follows: In Section \ref{secdos} we fix our notation, give some definitions, and establish the main assumption that must be considered throughout this work. In addition, we recall a characterization about the one-weight irreducible cyclic codes that will be useful. Section \ref{sectres} is devoted to recalling the Griesmer lower bound, and also to presenting three preliminary results. In Section \ref{seccuatro} we study a kind of exponential sums that help us to determine the weights, and their corresponding frequencies, of the codes in Theorem \ref{teouno}. In fact, we are going to present simple necessary and sufficient numerical conditions in order that the evaluation of an exponential sum of such kind is exactly equal to one. In Section \ref{seccinco} we use the definitions and results of the previous sections in order to present a formal proof of the Theorems \ref{teouno} and \ref{teodos}. After this, we will analyze the two easy-to-check necessary and sufficient numerical conditions of Theorem \ref{teodos} in order to give an explicit formula for the number of cyclic codes that satisfy such conditions. In addition we include, at the end of this section, some examples of Theorems \ref{teouno} and \ref{teodos}, as well as for such explicit formula. Finally, Section \ref{secseis} is devoted to presenting our conclusions.

\section{Notation, definitions, main assumption, and an already-known result}\label{secdos}

First of all we set for this section and the rest of this work, the following:

\medskip

\noindent
{\bf Notation.} By using $q$, $k$, $\Delta$, $e_1$ and $e_2$, we will denote five integers such that $q$ is the power of a prime number, $k$ is a positive integer, and $\Delta=(q^k-1)/(q-1)$. From now on, $\gamma$ will denote a fixed primitive element of $\bbbf_{q^k}$. For any integer $a$, the polynomial $h_a(x) \in \bbbf_{q}[x]$ will denote the minimal polynomial of $\gamma^{-a}$. Furthermore, we will denote by ``$\mbox{Tr}_{\bbbf_{q^k}/\bbbf_q}$" the trace mapping from $\bbbf_{q^k}$ to $\bbbf_q$. Lastly, by using $\chi'$ and $\chi$ we will denote, respectively, the canonical additive characters of $\bbbf_{q^k}$ and $\bbbf_{q}$.

A common integer operator in programming languages is the remainder, or modulus operator. This operator is commonly denoted as ``$\%$", and it is interesting to note that it is rarely used in mathematics, and this is so because the remainder of a division of two integers is commonly handled by means of the usual congruence relation among integer numbers. However, as we will see, this remainder operator will be especially important for this work, and therefore a formal definition of it is needed.

\begin{definition}
Let $A$ and $B$ be two integers such that $B>0$. Then, $A \% B$ (we read it as the {\em remainder of} $A$ {\em modulus} $B$), will represent the unique integer $r$ such that $0 \leq r < B$, and $r \equiv A \pmod{B}$.
\end{definition} 

As examples of the previous definition we have $9 \% 7=2$ and $(-9) \% 7=5$.

We, now set for this section and the rest of this work, the following:

\medskip

\noindent
{\bf Main assumption.} From now on, we are going to suppose that $\gcd(\Delta,e_2)=1$ (unless otherwise stated, $e_1$ is just any integer). Therefore, throughout all this work, we are going to reserve the Greek letters $\alpha$ and $\beta$ to represent any two integers such that $0 \leq \alpha < q^k-1$, $0 \leq \beta < q-1$, and $e_2\alpha+\Delta \beta \equiv 1 \pmod{q^k-1}$. In order to see that such pair of integers exists, assume that $S$ and $T$ are integers such that $e_2S+\Delta T = 1$. Then, we just need to take $\alpha=S\%(q^k-1)$ and $\beta=T\%(q-1)$.

An important type of irreducible cyclic codes are the so-called one-weight irreducible cyclic codes. The following is a characterization for them (see, for example, Theorem 2 in \cite{Vega2}): 

\begin{theorem}\label{teotres}
Let $a$ be any integer, and let $k'$, $u$, and $n$ be positive integers so that $u=\gcd(\frac{q^{k'}-1}{q-1},a)$, $\frac{q^{k'}-1}{\gcd(q^{k'}-1,a)} | n$, and $\deg(h_a(x))=k'$. Then, $h_a(x)$ is the parity-check polynomial of an $[n,k']$ one-weight irreducible cyclic code over $\bbbf_{q}$, whose nonzero weight is $\frac{n(q-1)}{q^{k'}-1} q^{k'-1}$ if and only if $u=1$.
\end{theorem}

\section{The Griesmer lower bound and some preliminary results}\label{sectres}

Let $n_q(k,d)$ be the minimum length $n$ for which an $[n,k,d]$ linear code, over $\bbbf_{q}$, exists. Given the values of $q$, $k$ and $d$, a central problem of coding theory is to determine the actual value of $n_q(k,d)$. A well-known lower bound (see \cite{Griesmer} and \cite{Solomon}) for $n_q(k,d)$ is

\begin{theorem}\label{teocuatro}
(Griesmer bound) With the previous notation,

\[n_q(k,d) \geq \sum_{i=0}^{k-1} \left \lceil \frac{d}{q^i} \right \rceil \; .\]
\end{theorem} 

With the aid of the previous lower bound, we now present the following:

\begin{lemma}\label{lemauno}
Suppose that ${\cal C}$ is a $[q^k-1,k+1,q^{k-1}(q-1)-1]$ linear code over $\bbbf_{q}$. Then ${\cal C}$ is an optimal linear code in the sense that its length reaches the lower bound in the previous theorem.
\end{lemma} 

\begin{proof} 
By means of a direct application of the Griesmer lower bound, we have

\begin{eqnarray}
&&\left \lceil \frac{q^{k-1}(q-1)-1}{q^0} \right \rceil + \left \lceil \frac{q^{k-1}(q-1)-1}{q} \right \rceil + \cdots + \left \lceil \frac{q^{k-1}(q-1)-1}{q^k} \right \rceil \nonumber \\
&=& [q^{k-1}(q-1)-1]+[q^{k-2}(q-1)]+\cdots+[q-1]+1=\Delta (q-1)=q^k-1 \; . \nonumber
\end{eqnarray}
\end{proof}

The following two results will be important in order to prove the characterization in Theorem \ref{teodos}.

\begin{lemma}\label{lemados}
Let ${\cal C}$ be a cyclic code of length $n$, over $\bbbf_{q}$, with parity-check polynomial $h(x)$. Suppose that ${\cal C}$ has exactly $q-1$ codewords of Hamming weight $n$. Then, there exists a unique polynomial $h'(x)$, of degree one, such that $h'(x) | h(x)$. 
\end{lemma}

\begin{proof} 
Let ${\cal M}=\{ \vec{c} \in {\cal C} \:|\: w_H(\vec{c})=n \}$, where $w_H(\cdot)$ stands for the usual Hamming weight function. Also let $\sigma: \bbbf_{q}^n \to \bbbf_{q}^n$ be the {\em circular shift} function defined by $\sigma(x_1,x_2,\cdots,x_n)=(x_2,\cdots,x_n,x_1)$, for all $(x_1,x_2,\cdots,x_n) \in \bbbf_{q}^n$. Now, let $\vec{m}=(m_1,m_2,\cdots,m_n)$ be a fixed element of ${\cal M}$. Since, $w_H(\lambda \vec{m})=n$, for all $\lambda \in \bbbf_{q}^*$, we can assume, without loss of generality, that $\vec{m}=(m_1=1,m_2,\cdots,m_n)$. Thus, 

\[{\cal M} \ni \sigma(\vec{m})=(m_2,\cdots,m_n,m_1)=(m_2 m_1,m_2 m_2,\cdots,m_2 m_n)=m_2 \vec{m} \in \{ \lambda \vec{m}  \:|\: \lambda \in \bbbf_{q}^* \}={\cal M} \; , \]

\noindent
therefore, $m_i=m_2^{i-1}$, for $i=1,2,\dots,n$, which in turn means that ${\cal M}=\{ (\lambda m_2^{i-1})_{i=1}^n \:|\: \lambda \in \bbbf_{q}^* \}$. That is, ${\cal C}':={\cal M} \cup \{\vec{0}\}$ is a cyclic code of length $n$ and dimension one, whose parity-check polynomial, $h'(x)$, is $h'(x)=x-m_2^{-1}$. But ${\cal C}' \subseteq {\cal C}$, hence $h'(x) | h(x)$.

Now, suppose that there exists another polynomial, $h''(x)$, of degree one, such that $h''(x) | h(x)$, and $h''(x) \neq h'(x)$. Under these circumstances, and thanks to Theorem \ref{teotres}, $h''(x)$ is the parity-check polynomial of another one-weight irreducible cyclic code of length $n$, and dimension one, whose nonzero weight is $n$. Therefore, ${\cal C}$ must have at least $2(q-1)$ codewords of Hamming weight $n$. A contradiction!  
\end{proof}

The following lemma can be seen as an almost direct consequence of the work of Schmidt and White in \cite{Schmidt} (see particularly Corollary 3.2 therein, and Theorem 6 in \cite{Vega2}). As will be clear later, this lemma will be one of the main arguments that will make possible to prove our extended characterization.

\begin{lemma}\label{lematres}
Let $k'$ be a divisor $k$, and for any suitable integer $e$, let ${\cal C}_{(e)}$ be a $[q^k-1,k']$ two-weight irreducible cyclic code, over $\bbbf_q$, whose parity check polynomial is $h_e(x)$. Suppose that $q=p^t$, for some integer $t$, and some prime $p$. Thus, if $w_1$ and $w_2$ are the nonzero weights of ${\cal C}_{(e)}$, then $w_1-w_2 \neq \pm 1$.
\end{lemma}

\begin{proof}
If $k'<k$, then $\frac{q^k-1}{q^{k'}-1}>1$, and since $\frac{q^k-1}{q^{k'}-1} | w_i$, for $i=1,2$, $w_1-w_2 \neq \pm 1$. Suppose $k'=k$. Thus, for a positive integer $x$, let $S_p(x)$ denote the sum of the $p$-digits of $x$. Then, since ${\cal C}_{(e)}$ is a $[q^k-1,k]$ two-weight irreducible cyclic code, we have, owing to Theorem 6 in \cite{Vega2}, that $w_1=\frac{q-1}{q}(q^k - r \varepsilon p^{s\theta})$ and $w_2=\frac{q-1}{q}(q^k + (u-r) \varepsilon p^{s\theta})$, where $u=\gcd(\Delta,e)>1$, $f:=\mbox{ord}_{u}(p)$, $kt=fs$, $\varepsilon=\pm 1$,

$$\theta=\frac{1}{p-1} \min \left\{ S_p\left(\frac{j(p^f-1)}{u}\right) \; | \; 1 \leq j < u \right\} \;, $$

\noindent
and $r$ is a positive integer satisfying: 

\begin{eqnarray}
&& r | (u-1) \;, \nonumber \\
&& r p^{s\theta} \equiv \pm 1 \pmod{u} \;, \nonumber \\
&& r (u-r)=(u-1) p^{s(f-2\theta)} \; . \nonumber 
\end{eqnarray}

Thus, we have $w_1-w_2 = \pm \frac{q-1}{q} u p^{s\theta}$. But $u>1$, and $\gcd(q,u)=1$, therefore $w_1-w_2 \neq \pm 1$.
\end{proof}

\section{A class of exponential sums}\label{seccuatro}

We want to recall that $\alpha$ and $\beta$ are integers, with $0 \leq \alpha < q^k-1$ and $0 \leq \beta < q-1$, such that $e_2\alpha+\Delta \beta \equiv 1 \pmod{q^k-1}$ (see Main assumption). Now, let $i$, $j$ and $v$ be integers such that $0 \leq i < q^k-1$, $0 \leq j < q-1$ and $v=(e_2i+\Delta j)\%(q^k-1)$. Then, note that the previous equality implies that $e_2 \alpha v + \Delta \beta v \equiv e_2 i + \Delta j \equiv v \pmod{q^k-1}$, which in turn implies that $\Delta (\beta v-j) \equiv e_2(i-\alpha v) \pmod{q^k-1}$. But this last congruence has solution if and only if $\Delta | (i-\alpha v)$ (see, for example, Proposition 3.3.1 in \cite{Ireland}), that is, if and only if $i=(\alpha v + \Delta w)\%(q^k-1)$ and $j=(\beta v - e_2 w)\%(q-1)$, where $w=(\frac{i-\alpha v}{\Delta})\%(q-1)$. Thus, by keeping this in mind, we now present the following:

\begin{lemma}\label{lemacuatro}
Let 

\[{\cal V} := \{(i,j) \: | \: 0 \leq i < q^k-1 \mbox{ and } 0 \leq j < q-1 \} \: .\]

\noindent
Consider the map $\Phi: {\cal V} \to {\cal V}$, given by the rule $\Phi(i,j)=(v,w)$, where $v=(e_2i+\Delta j)\%(q^k-1)$ and $w=(\frac{i-\alpha v}{\Delta})\%(q-1)$. Then $\Phi$ is a bijective map over ${\cal V}$.
\end{lemma} 

\begin{proof} 
Let $(i,j), (i',j') \in {\cal V}$ such that $\Phi(i,j)=\Phi(i',j')=(v,w)$. Then, $\frac{i-\alpha v}{\Delta} \equiv \frac{i'-\alpha v}{\Delta} \equiv w \pmod{q-1}$, or equivalently, $i-\alpha v \equiv i'-\alpha v \equiv \Delta w \pmod{q^k-1}$. But $0 \leq i,i' < q^k-1$, therefore $i=i'$. In a similar way, $e_2i+\Delta j \equiv e_2i'+\Delta j' \equiv v \pmod{q^k-1}$. But we already know that $i=i'$, and since $q^k-1=\Delta(q-1)$ and $0 \leq j,j' < q-1$, we can conclude that $j=j'$. Thus, $\Phi$ is injective. Now, for $(v,w) \in {\cal V}$ we take $i=(\alpha v + \Delta w)\%(q^k-1)$ and $j=(\beta v - e_2 w)\%(q-1)$. For such a choice of the pair $(i,j)$, we have $(e_2i+\Delta j)\%(q^k-1)=(e_2\alpha v+e_2\Delta w+\Delta \beta v - \Delta e_2w)\%(q^k-1)=((e_2\alpha+\Delta \beta)v)\%(q^k-1)=v$, and $(\frac{i-\alpha v}{\Delta})\%(q-1)=(\frac{\alpha v + \Delta w -\alpha v}{\Delta})\%(q-1)=w$. Thus, $\Phi$ is also surjective, and $\Phi$ is a bijective map.
 \end{proof}

\begin{lemma}\label{lemacinco}
With our notation, let $d:=\gcd(q-1,ke_1-e_2)$. Thus, if $d>1$, then

\[\Delta (e_1 \alpha+\beta) \equiv 1 \pmod{d} \; .\]
\end{lemma}

\begin{proof} 
Since $d | (q^k-1)$, $e_2\alpha+\Delta \beta \equiv 1 \pmod{d}$. On the other hand, since $\Delta \equiv k \pmod{q-1}$ (see, for example, Remark 3 in \cite{Vega1}), $d=\gcd(q-1,\Delta e_1-e_2)$, and consequently $d | (\Delta e_1 - e_2)\alpha$. Therefore, $\Delta (e_1 \alpha+\beta) \equiv e_2\alpha+\Delta \beta \equiv 1  \pmod{d}$.
\end{proof}

\begin{lemma}\label{lemaseis}
With the same notation as above, let $(a,b) \in \bbbf_{q^k}^2$. Define:

\begin{eqnarray}
f_{a,b,d} : {\cal V} &\to& \bbbf_{q^k} \; , \nonumber \\ 
f_{a,b,d}(v,w) &:=& a \gamma^{\Delta(e_1\alpha+\beta)v+\Delta d w} + b \gamma^{v}\; . \nonumber
\end{eqnarray}

\noindent
If $\rho := \frac{\Delta(e_1 \alpha + \beta)-1}{d}$, then, for any integer $r$, and for $t=0,1,\dots,d-1$, we have:

\begin{eqnarray}\label{vequno}
\gamma^{r\Delta}f_{a,b,d}(v,w)&=&f_{a,b,d}((v+r\Delta)\%(q^k-1),(w-r\rho)\%(q-1)) \: ,  \\
f_{a,b,d}(v,w)&=&f_{a,b,d}(v,(w+\frac{q-1}{d}t)\%(q-1)) \; . \label{veqdos}
\end{eqnarray}
\end{lemma}

\begin{proof} 
The proof is almost direct from the definition of $f_{a,b,d}(v,w)$;

\begin{eqnarray}
\gamma^{r\Delta}f_{a,b,d}(v,w)&=&a \gamma^{\Delta(e_1\alpha+\beta)v + d \Delta w + r\Delta}+b\gamma^{v+r\Delta}\nonumber \\ 
&=&a \gamma^{\Delta(e_1\alpha+\beta)(v + r\Delta) + d \Delta w +(1-\Delta(e_1 \alpha + \beta))r\Delta}+b\gamma^{v+r\Delta}\nonumber \\ 
&=&a \gamma^{\Delta(e_1\alpha+\beta)(v + r\Delta) + d \Delta (w -r\rho)}+b\gamma^{v+r\Delta}\nonumber \\ 
&=&f_{a,b,d}((v+r\Delta)\%(q^k-1),(w-r\rho)\%(q-1)) \nonumber \; .
\end{eqnarray}

On the other hand,

\begin{eqnarray}
f_{a,b,d}(v,(w+\frac{q-1}{d}t)\%(q-1))&=&a \gamma^{\Delta(e_1\alpha+\beta)v + d \Delta w + \Delta(q-1)t}+b\gamma^{v}\nonumber \\ 
&=&a \gamma^{\Delta(e_1\alpha+\beta)v + d \Delta w}+b\gamma^{v}=f_{a,b,d}(v,w) \nonumber \; . 
\end{eqnarray}
\end{proof}

\begin{remark}\label{rmuno} 
Through the function $f_{a,b,d}(v,w)$ we induce a disjoint partition of ${\cal V}$, as follows:

\begin{eqnarray}
{\cal V}_{0}&=& \{(v,w) \in {\cal V} \: | \: f_{a,b,d}(v,w)=0 \} \: , \mbox{ and }  \nonumber \\ 
{\cal V}_{\gamma^i}&=&\{(v,w) \in {\cal V} \: | \: f_{a,b,d}(v,w)=\gamma^i \} \nonumber \; , 
\end{eqnarray}

\noindent
for $i=0,1,\dots,q^k-2$. Clearly, these subsets are disjoint and ${\cal V}={\cal V}_{0} \cup (\cup_{i=0}^{q^k-2} {\cal V}_{\gamma^i})$. In addition, by (\ref{vequno}) we have $|{\cal V}_{\gamma^{i + r\Delta}}|=|{\cal V}_{\gamma^{i}}|$ for any integer $r$. On the other hand, by (\ref{veqdos}) we also have that there must exist $q^k$ non-negative integers, $N,N_0,N_1,\dots,N_{q^k-2}$, such that $|{\cal V}_{0}|=dN$ and $|{\cal V}_{\gamma^{i}}|=dN_{i}$. By combining $|{\cal V}_{\gamma^{i + r\Delta}}|=|{\cal V}_{\gamma^{i}}|$, with $|{\cal V}_{\gamma^{i}}|=dN_{i}$, we get $N_{(i + r\Delta)\%(q^k-1)}=N_{i\%(q^k-1)}$, for all integers $i$ and $r$.
\end{remark}

The following result is a generalization of the characterization given in Lemma 3 of \cite{Vega3}, that shows that Lemma 6, in \cite[p. 4503]{Heng}, can be upgraded to a characterization. Now, note that an interesting feature of the following result is that its proof relies completely on the remainder operator, and on the previous three lemmas, which, by the way, are also based on such operator. Therefore, as we will see next, thanks to this feature we are able to present a simplified and self-contained proof of this generalized characterization.

\begin{lemma}\label{lemasiete}
Let $(a,b) \in \bbbf_{q^k}^2$, and suppose $\mbox{Tr}_{\bbbf_{q^k}/\bbbf_{q}}(a) \neq 0$ and $b \neq 0$. Consider the sums of the form:

\[T_{(e_1,e_2)}(a,b) := \sum_{x \in \bbbf_{q^k}^*} \sum_{y \in \bbbf_{q}^*} \chi'(a x^{\Delta e_1}y + b x^{e_2}y) \; .\]

\noindent
Then $\gcd(q-1,ke_1-e_2)=1$ if and only if $T_{(e_1,e_2)}(a,b)=1$.
\end{lemma}

\begin{proof} 
Clearly,

\[T_{(e_1,e_2)}(a,b) = \sum_{i=0}^{q^k-2} \sum_{j=0}^{q-2} \chi'(a\gamma^{\Delta e_1i}\gamma^{\Delta j} + b \gamma^{e_2i}\gamma^{\Delta j}) \; ,\] 

\noindent
and owing to Lemma \ref{lemacuatro}, we can apply the variable substitutions $i \mapsto (\alpha v + \Delta w)\%(q^k-1)$, and $j \mapsto (\beta v - e_2 w)\%(q-1)$ (recall that $e_2\alpha+\Delta \beta \equiv 1 \pmod{q^k-1}$). Thus,

\begin{eqnarray}\label{lequno}
T_{(e_1,e_2)}(a,b) &=& \sum_{v=0}^{q^k-2} \sum_{w=0}^{q-2} \chi'(a\gamma^{\Delta(e_1\alpha+\beta)v} \gamma^{\Delta(\Delta e_1-e_2)w} + b\gamma^{v}) \nonumber \\ 
&=& \sum_{v=0}^{q^k-2} \sum_{w=0}^{q-2} \chi'(a\gamma^{\Delta(e_1\alpha+\beta)v} \gamma^{\Delta(k e_1-e_2)w} + b\gamma^{v})  \; , 
\end{eqnarray}

\noindent
because $\Delta e_1-e_2 \equiv k e_1-e_2 \pmod{q-1}$. Now, if $\gcd(q-1,k e_1-e_2)=1$, then

\[T_{(e_1,e_2)}(a,b)=\sum_{v=0}^{q^k-2} \sum_{w=0}^{q-2} \chi'(a\gamma^{\Delta w} + b\gamma^{v})=\sum_{y \in \bbbf_{q}^*} \sum_{x \in \bbbf_{q^k}^*} \chi'(a y + b x) \; ,\] 

\noindent
and since $b \neq 0$, we have

\begin{eqnarray}
T_{(e_1,e_2)}(a,b)&=& \sum_{y \in \bbbf_{q}^*} \sum_{x \in \bbbf_{q^k} \setminus \{ay\}} \chi'(x) \nonumber \\
&=& \sum_{y \in \bbbf_{q}^*} (\sum_{x \in \bbbf_{q^k}} \chi'(x) - \chi'(a y)) \nonumber \\
&=&-\sum_{y \in \bbbf_{q}^*} \chi'(a y) = - \sum_{y \in \bbbf_{q}^*} \chi(\mbox{Tr}_{\bbbf_{q^k}/\bbbf_{q}}(a)y)=1 \; , \nonumber
\end{eqnarray}

\noindent
because $\mbox{Tr}_{\bbbf_{q^k}/\bbbf_{q}}(a) \neq 0$. 

On the other hand, if $\gcd(q-1,k e_1-e_2)=d>1$ then, from (\ref{lequno}), we have

\begin{eqnarray}
T_{(e_1,e_2)}(a,b) &=& \sum_{v=0}^{q^k-2} \sum_{w=0}^{q-2} \chi'(a\gamma^{\Delta(e_1\alpha+\beta)v} \gamma^{\Delta d w} + b\gamma^{v}) \nonumber \\
&=& \sum_{v=0}^{q^k-2} \sum_{w=0}^{q-2} \chi'(f_{a,b,d}(v,w)) \; , \nonumber
\end{eqnarray}

\noindent
where the last equality arises from the definition of the function $f_{a,b,d}(v,w)$ in Lemma \ref{lemaseis}. Now, by considering the discussion and the notation of Remark \ref{rmuno}, we have 

\[T_{(e_1,e_2)}(a,b)=|{\cal V}_{0}| \chi'(0) + \sum_{i=0}^{q^k-2} |{\cal V}_{\gamma^{i}}| \chi'(\gamma^{i})=d(N + \sum_{i=0}^{q^k-2} N_{i} \chi'(\gamma^{i})) \; , \]

\noindent
however, as was pointed out in Remark \ref{rmuno}, $N_{(i + r\Delta)\%(q^k-1)}=N_{i\%(q^k-1)}$ for all integers $i$ and $r$, therefore,

\begin{eqnarray}
T_{(e_1,e_2)}(a,b) &=& dN + d\sum_{i=0}^{\Delta-1} N_{i} \sum_{r=0}^{q-2} \chi'(\gamma^{i + r\Delta}) \nonumber \\
&=& dN + d\sum_{i=0}^{\Delta-1} N_{i} \sum_{y \in \bbbf_{q}^*} \chi(\mbox{Tr}_{\bbbf_{q^k}/\bbbf_{q}}(\gamma^{i})y) \; . \nonumber
\end{eqnarray}

\noindent
Now, let ${\cal I}=\{0,1,\cdots,\Delta-1\}$, and, by considering the subset ${\cal I}_{0}=\{ i \in  {\cal I} \:|\: \mbox{Tr}_{\bbbf_{q^k}/\bbbf_{q}}(\gamma^{i})=0 \}$, we have

\begin{eqnarray}
T_{(e_1,e_2)}(a,b) &=& dN + d\sum_{i \in {\cal I}_{0}} (q-1)N_{i} \chi(0) + d\sum_{i \in {\cal I} \setminus {\cal I}_{0}} N_{i} \sum_{y \in \bbbf_{q}^*} \chi(\mbox{Tr}_{\bbbf_{q^k}/\bbbf_{q}}(\gamma^{i})y) \nonumber \\
&=& d(N + (q-1)\sum_{i \in {\cal I}_{0}} N_{i} - \sum_{i \in {\cal I} \setminus {\cal I}_{0}} N_{i})=dt \; , \nonumber
\end{eqnarray}

\noindent
for some integer $t$, and since $d>1$, we have $T_{(e_1,e_2)}(a,b) \neq 1$.
\end{proof}
 
We end this section with the following:

\begin{corollary}\label{coruno}
Assume the same notation as before. Thus, if $\gcd(q-1,ke_1-e_2)=1$ and $\gcd(\Delta,e_2)=1$, then

\[T_{(e_1,e_2)}(a,b)=
\left\{ \begin{array}{cl}
		(q-1)(q^k-1) & \mbox{ if $\;\;\;a=0$ and $b=0$,} \\
		-(q^k-1)     & \mbox{ if $\mbox{Tr}_{\bbbf_{q^k}/\bbbf_{q}}(a) \neq 0$ and $b=0$,} \\
		-(q-1)       & \mbox{ if $\mbox{Tr}_{\bbbf_{q^k}/\bbbf_{q}}(a) =0$ and $b \neq 0$,} \\
		 1           & \mbox{ if $\mbox{Tr}_{\bbbf_{q^k}/\bbbf_{q}}(a) \neq 0$ and $b \neq 0$}.
			\end{array}
\right . \]
\end{corollary}

\begin{proof}
Clearly, $T_{(e_1,e_2)}(0,0)=(q-1)(q^k-1)$, and if $\mbox{Tr}_{\bbbf_{q^k}/\bbbf_{q}}(a) \neq 0$ and $b=0$, then 

\begin{eqnarray}
T_{(e_1,e_2)}(a,0)&=&\sum_{y \in \bbbf_{q}^*} \sum_{x \in \bbbf_{q^k}^*}\chi'(a x^{\Delta e_1}y) \nonumber \\
&=&\sum_{x \in \bbbf_{q^k}^*} \sum_{y \in \bbbf_{q}^*}\chi(\mbox{Tr}_{\bbbf_{q^k}/\bbbf_{q}}(a) x^{\Delta e_1}y)=-(q^k-1) \; . \nonumber
\end{eqnarray}

\noindent
On the other hand, if $\mbox{Tr}_{\bbbf_{q^k}/\bbbf_{q}}(a)=0$ and $b \neq 0$, then 

\begin{eqnarray}
T_{(e_1,e_2)}(a,b)&=&\sum_{y \in \bbbf_{q}^*} \sum_{x \in \bbbf_{q^k}^*} \chi(\mbox{Tr}_{\bbbf_{q^k}/\bbbf_{q}}(a) x^{\Delta e_1}y)\chi'(b x^{e_2}y) \nonumber \\
&=&\sum_{x \in \bbbf_{q^k}^*} \sum_{y \in \bbbf_{q}^*} \chi(\mbox{Tr}_{\bbbf_{q^k}/\bbbf_{q}}(b x^{e_2})y) \;,  \nonumber 
\end{eqnarray}

\noindent
but, since $\gcd(\Delta,e_2)=1$, we have that $|\{ x \in \bbbf_{q^k} \:|\: \mbox{Tr}_{\bbbf_{q^k}/\bbbf_{q}}(b x^{e_2})=0 \}|=q^{k-1}$. Therefore

\[T_{(e_1,e_2)}(a,b)=(q^{k-1}-1)(q-1) - q^{k-1}(q-1)=-(q-1) \;.\]

\noindent
Finally, the proof of the last case comes from the previous lemma.
\end{proof}

\section{Formal proof of Theorems \ref{teouno} and \ref{teodos}}\label{seccinco}

We can now present a formal proof of Theorem \ref{teouno}.

\begin{proof}
Part {(A)}: Since $\Delta e_1 q \equiv \Delta e_1 \pmod{q^k-1}$, $\deg(h_{\Delta e_1}(x))=1$. Let $l$ be the smallest positive integer such that $e_2 q^l \equiv e_2 \pmod{q^k-1}$. Thus $\Delta | e_2\frac{q^l-1}{q-1}$. But $\gcd(\Delta,e_2)=1$, therefore $(q^k-1) | (q^l-1)$, which in turn implies that $\deg(h_{\Delta e_1}(x))=l=k$. Now, owing to Theorem \ref{teotres}, we know that ${\cal C}_{(\Delta e_1)}$ is a one-weight irreducible cyclic code, whose nonzero weight is $q^k-1$. On the other hand, because $\gcd(\Delta,e_2)=1$, we can conclude, in a similar manner, that $h_{e_2}(x)$ is the parity-check polynomial of a one-weight cyclic code of length $q^k-1$, whose nonzero weight is $q^{k-1}(q-1)$.

Part {(B)}: ${\cal C}_{(\Delta e_1,e_2)}$ is a $[q^k-1,k+1]$ cyclic code due to Part {(A)}. Let ${\cal A}$ be a fixed subset of $\bbbf_{q^k}^*$ so that $\{\mbox{Tr}_{\bbbf_{q^k}/\bbbf_q}(a) \:|\: a \in {\cal A}\}=\bbbf_{q}^*$. Now, for each $a \in {\cal A} \cup \{0\}$ and $ b \in \bbbf_{q^k}$, we define $c(q^k-1,e_1,e_2,a,b)$ as the vector of length $q^k-1$ over $\bbbf_q$, which is given by:

$$(\mbox{Tr}_{\bbbf_{q^k}/\bbbf_q}(a(\gamma^{\Delta e_1})^i+b(\gamma^{e_2})^i))_{i=0}^{q^k-2}\; .$$

\noindent
Thanks to Delsarte's Theorem (see, for example, \cite{Delsarte}) it is well known that 

$${\cal C}_{(\Delta e_1,e_2)}=\{ c(q^k-1,e_1,e_2,a,b) \: | \: a \in {\cal A} \cup \{0\}, \mbox{ and } b \in \bbbf_{q^k} \} \; .$$ 

\noindent
Thus the Hamming weight of any codeword $c(q^k-1,e_1,e_2,a,b)$, will be equal to $q^k-1-Z(a,b)$, where $Z(a,b)\!=\!\sharp\{\;i\; | \; \mbox{Tr}_{\bbbf_{q^k}/\bbbf_q}(a\gamma^{\Delta e_1 i}+b\gamma^{e_2 i})=0, \: 0 \leq i < q^k-1 \}$. That is, we have

\begin{eqnarray}
Z(a,b)&=&\frac{1}{q}\sum_{i=0}^{q^k-2} \sum_{y \in \bbbf_q} \chi(\mbox{Tr}_{\bbbf_{q^k}/\bbbf_q}((a\gamma^{\Delta e_1 i}+b\gamma^{e_2 i})y)) \nonumber \\
&=&\frac{q^k-1}{q}+\frac{1}{q}\sum_{y \in \bbbf_q^*} \sum_{x \in \bbbf_{q^k}^*} \chi'(a x^{\Delta e_1}y+b x^{e_2}y)  \; , \nonumber
\end{eqnarray}   

\noindent
and, by using the notation of Lemma \ref{lemasiete}, we have

\begin{equation}\label{eqcinco}
Z(a,b)=\frac{q^k-1}{q}+\frac{1}{q} T_{(e_1,e_2)}(a,b) \; .
\end{equation}

\noindent
But $\gcd(q-1,ke_1-e_2)=1$ and $\gcd(\Delta,e_2)=1$; therefore, after applying Corollary \ref{coruno}, we get

\[Z(a,b)=\left\{ \begin{array}{cl}
		q^k-1 & \mbox{ if $a=0$ and $b=0$,} \\
		0    & \mbox{ if $a \in {\cal A}$ and $b=0$,} \\
		q-1      & \mbox{ if $a=0$ and $b \neq 0$,} \\
		q       & \mbox{ if $a \in {\cal A}$ and $b \neq 0$}.
			\end{array}
\right . \]

\noindent
Consequently, the assertion about the weight distribution of ${\cal C}_{(\Delta e_1,e_2)}$ comes now from the fact that the Hamming weight of any codeword in ${\cal C}_{(\Delta e_1,e_2)}$ is equal to $q^k-1-Z(a,b)$, and also due to the fact that $|{\cal A}|=q-1$ and $|\bbbf_{q^k}^*|=q^k-1$.  

Lastly, ${\cal C}_{(\Delta e_1,e_2)}$ is an optimal cyclic code, due to Lemma \ref{lemauno}, and the assertion about the weights of the dual code of ${\cal C}_{(\Delta e_1,e_2)}$ can now be proved by means of Table I and the first four identities of Pless (see, for example, pp. 259-260 in \cite{Huffman}).
\end{proof}

We continue by presenting now a formal proof of Theorem \ref{teodos}.

\begin{proof}
Suppose that ${\cal C}$ is a cyclic code of length $q^k-1$, over $\bbbf_{q}$, whose weight distribution is given in Table I. Through the sum of the frequencies of such table, it is easy to see that ${\cal C}$ must be a cyclic code of dimension $k+1$. Consequently, the degree of the parity-check polynomial $h(x)$, of ${\cal C}$, must be equal to $k+1$. Now, note that for any integer $e$ we have that $\deg(h_{\Delta e}(x))=1$, therefore, thanks to Lemma \ref{lemados}, there must exist an integer $e_1$ such that $h_{\Delta e_1}(x) | h(x)$. Let $h'(x)\neq 1$ be an irreducible divisor of $h(x)/h_{\Delta e_1}(x)$, thus, if $k'=\deg(h'(x))$, then $k'>1$ (owing to Lemma \ref{lemados}), and $k' | k$. Also, let ${\cal C}'$ be the irreducible cyclic code of length $q^k-1$, over $\bbbf_{q}$, whose parity-check polynomial is $h'(x)$. Since ${\cal C}' \subsetneq {\cal C}$, the cyclic code ${\cal C}'$ has at most two nonzero weights, and, in accordance with Table I, these nonzero weights may only be $w_1:=q^{k-1}(q-1)-1$ and $w_2:=q^{k-1}(q-1)$. Thus, owing to Lemma \ref{lematres}, and since $w_1-w_2=-1$, ${\cal C}'$ cannot be a two-weight irreducible cyclic code. Suppose then that ${\cal C}'$ is a one-weight irreducible cyclic code of length $q^k-1$. Now, by further supposing that $k'<k$, we obtain, thanks to Theorem \ref{teotres}, that the nonzero weight of ${\cal C}'$ is $\frac{q^k-1}{q^{k'}-1}(q-1)q^{k'-1} > (q-1)q^{k-1}$. Therefore, this nonzero weight cannot be equal to either $w_1$ or $w_2$. In consequence, $h'(x)=h(x)/h_{\Delta e_1}(x)$ is the parity-check polynomial of a $[q^k-1,k]$ one-weight irreducible cyclic code, whose nonzero weight $q^{k-1}(q-1)$. But by considering again Theorem \ref{teotres}, the previous fact implies that there must exist an integer $e_2$ such that $\gcd(\Delta, e_2)=1$, and $h(x)=h_{\Delta e_1}(x)h_{e_2}(x)$. 

It remains to prove that $\gcd(q-1,k e_1-e_2)=1$. Let ${\cal C}_{(\Delta e_1)}$ and ${\cal C}_{(e_2)}$ be the irreducible cyclic codes of length $q^k-1$ over $\bbbf_{q}$, whose parity-check polynomials are, respectively, $h_{\Delta e_1}(x)$ and $h_{e_2}(x)$. If $\gcd(\Delta,e_2)=1$, then, once again, ${\cal C}_{(e_2)}$ will correspond to a one-weight irreducible cyclic code of length $q^k-1$ and dimension $k$, whose nonzero weight is $q^k(q-1)$. Since, in Table I, the frequency of such nonzero weight is $q^k-1=|\bbbf_{q^k}^*|$ we have that a codeword $c$, in ${\cal C}$, will have Hamming weight $q^k(q-1)-1$ if and only if $c=c_1+c_2$, where $c_1$ and $c_2$ are, respectively, two nonzero codewords in ${\cal C}_{(\Delta e_1)}$ and ${\cal C}_{(e_2)}$. But if $c_1$ and $c_2$ are nonzero codewords in ${\cal C}_{(\Delta e_1)}$ and ${\cal C}_{(e_2)}$, then there must exist two finite field elements $a$ and $b$ in $\bbbf_{q^k}$, with $\mbox{Tr}_{\bbbf_{q^k}/\bbbf_{q}}(a) \neq 0$, $b \neq 0$, so that the number of zero entries, $Z(a,b)$, in codeword $c$, can be computed by means of (\ref{eqcinco}). Under these circumstances, codeword $c$ will have Hamming weight $q^k(q-1)-1$ if and only if $T_{(e_1,e_2)}(a,b)=1$, and due to Lemma \ref{lemasiete}, this can only be possible if and only if $\gcd(q-1,ke_1-e_2)=1$. 

Finally, the proof of the converse is just a part of the proof of Theorem \ref{teouno} that was already given.
\end{proof}

Due to the simplicity of the necessary and sufficient numerical conditions in Theorem \ref{teodos}, it is possible to compute the total number of different cyclic codes, over $\bbbf_{q}$, of length $q^k-1$ and dimension $k$, that satisfy such conditions. The following result goes in that direction.

\begin{theorem}\label{teocinco}
With our notation, let ${\cal N}$ be the number of different cyclic codes, ${\cal C}_{(\Delta e_1,e_2)}$, of length $q^k-1$ and dimension $k+1$ that satisfy conditions in Theorem \ref{teodos}. Then

\begin{equation}\label{eqseis}
{\cal N} = \frac{\phi(q^k-1)(q-1)}{k} \; ,
\end{equation}

\noindent
where $\phi$ denotes the Euler $\phi$-function.
\end{theorem} 

\begin{proof}
Since $\deg(h_{e_2}(x))=k$, the total number, ${\cal N}_2$, of different minimal polynomials $h_{e_2}(x)$ that satisfy condition $\gcd(\Delta,e_2)=1$ is ${\cal N}_2=\frac{\phi(\Delta)(q-1)}{k}$. On the other hand, since $\deg(h_{\Delta e_1}(x))=1$ we have that for each integer $e_2$ that satisfies $\gcd(\Delta,e_2)=1$, the total number, ${\cal N}_1$, of different minimal polynomials $h_{\Delta e_1}(x)$ that satisfy condition $\gcd(q-1,ke_1-e_2)=1$ is ${\cal N}_1=\phi(q-1)\frac{d}{\phi(d)}$, where $d=\gcd(k,q-1)$. Now, recall that for any two positive integers $m$ and $n$, we have $\phi(m n)=\phi(m)\phi(n)\frac{\delta}{\phi(\delta)}$, where $\delta=\gcd(m,n)$. Thus, $\phi(q^k-1)=\phi(\Delta (q-1))=\phi(\Delta)\phi(q-1)\frac{d'}{\phi(d')}$, where $d'=\gcd(\Delta,q-1)$. In consequence, the result follows from the fact that ${\cal N}={\cal N}_1 {\cal N}_2$ and $d'=d$.
\end{proof}

As a direct consequence of the previous theorem and Theorems \ref{teodos}, we have the following:

\begin{corollary}\label{cordos}
Let ${\cal N}$ be the number of different cyclic codes of length $q^k-1$, over $\bbbf_{q}$, whose weight distribution is given in Table I. Then ${\cal N}$ is given by (\ref{eqseis}).
\end{corollary}

The following are examples related to Theorems \ref{teouno} and \ref{teodos}, and Corollary \ref{cordos}.


\begin{example}\label{ejecero}
With our notation, let $q=4$, $k=3$, $e_1=2$ and $e_2=5$. Then $\Delta=21$, $\gcd(21,e_2)=1$ and $\gcd(q-1,3e_1-e_2)=1$. Therefore, by Theorem \ref{teouno}, we can be sure that ${\cal C}_{(42,5)}$ is an optimal three-weight cyclic code over $\bbbf_{4}$, of length 63, dimension 4 and weight enumerator polynomial: $1+189z^{47}+63z^{48}+3z^{63}$. In addition, $B_1=B_2=0$, and $B_3=3843$. In fact, the dual code of ${\cal C}_{(42,5)}$ is a $[63,59,3]$ cyclic code over $\bbbf_{4}$ which, by the way, has the same parameters as the best known linear code, according to the tables of the best known linear codes maintained by Markus Grassl at http://www.codetables.de/.
\end{example}

\begin{example}\label{ejeuno}
With our notation, let $q=3$ and $k=4$. Then, owing to Corollary \ref{cordos}, the total number of different cyclic codes of length $80$, over $\bbbf_{3}$, and dimension $5$, with weight enumerator polynomial $1+160z^{53}+80z^{54}+2z^{80}$, is ${\cal N}=16$. In fact, these cyclic codes are: ${\cal C}_{(0,1)}$, ${\cal C}_{(0,7)}$, ${\cal C}_{(0,11)}$, ${\cal C}_{(0,13)}$, ${\cal C}_{(0,17)}$, ${\cal C}_{(0,23)}$, ${\cal C}_{(0,41)}$, ${\cal C}_{(0,53)}$, ${\cal C}_{(40,1)}$, ${\cal C}_{(40,7)}$, ${\cal C}_{(40,11)}$, ${\cal C}_{(40,13)}$, ${\cal C}_{(40,17)}$, ${\cal C}_{(40,23)}$, ${\cal C}_{(40,41)}$, and ${\cal C}_{(40,53)}$. 
\end{example}

\section{Conclusions}\label{secseis}

As we already mentioned, in coding theory the weight distribution problem of a cyclic code is an important issue. However, most of the conventional methods employed for the weight distribution computations require the use of, for example, Gauss and/or Jacobi sums along with very sophisticated --but at the same time complex-- theorems (for example the Davenport-Hasse Theorem). In this work we used some new and non-conventional methods in order to extend a characterization for the weight distribution of a class of three-weight cyclic codes of dimension 3, to a characterization for the weight distribution of a class of three-weight cyclic codes of dimension greater than or equal to 3, that includes the first characterized class. More specifically, we used the remainder operator, which is quite common in programming languages, in order to show that the numerical conditions given in Theorem 11 of \cite[p. 4505]{Heng} are also necessary, and as a consequence of this, we were able to upgrade such theorem to an extended characterization (Theorems \ref{teouno} and \ref{teodos}) that includes the characterization given in \cite{Vega3}. Furthermore, we would like to emphasize that by using the remainder operator, we were also able to present a simplified and self-contained proof of our extended characterization. Finally, we also found the parameters for the dual code of any cyclic code in our extended characterization class, and after the analysis of some examples, it seems that such dual codes always have the same parameters as the best known linear codes.

\bibliographystyle{IEEE}

\begin{thebibliography}{1}
\bibitem{Delsarte} P. Delsarte, On subfield subcodes of Reed-Solomon codes, IEEE Trans. Inf. Theory, IT-21(5) (1975) 575-576.

\bibitem{Griesmer} J.H. Griesmer, A bound for error correcting codes, IBM J. Res. Dev. 4 (1960) 532-542.

\bibitem{Heng} Z. Heng and Q. Yue, Several Classes of Cyclic Codes With Either Optimal Three Weights or a Few Weights, IEEE Trans. Inf. Theory, vol. 62(8) (2016) 4501-4513.

\bibitem{Huffman} W.C. Huffman and V.S Pless, Fundamental of Error-Correcting Codes, Cambridge Univ. Press, Cambridge, 2003.

\bibitem{Ireland} K. Ireland and M. Rosen, A Classical Introduction to Modern Number Theory, Springer-Verlag, 1982.

\bibitem{Schmidt} B. Schmidt and C. White, All two-weight irreducible cyclic codes?, Finite Fields and their Appl. 8 (2002) 1-17.

\bibitem{Solomon} G. Solomon and J.J. Stiffler, Algebraically punctured cyclic codes, Inform. and Control 8 (1965) 170-179.

\bibitem{Vega1} G. Vega, The Weight Distribution of an Extended Class of Reducible Cyclic Codes, IEEE Trans. Inf. Theory, vol. 58(7) (2012) 4862-4869.

\bibitem{Vega2} G. Vega, A critical review and some remarks about one- and two-weight irreducible cyclic codes, Finite Fields Appl. 33 (2015) 1-13.

\bibitem{Vega3} G. Vega, A characterization of a class of optimal three-weight cyclic codes of dimension 3 over any finite field, Finite Fields Appl. 42 (2016) 23-38.
\end{thebibliography}

\end{document}